\documentclass[preprint,showpacs,preprintnumbers,amsmath,amssymb]{revtex4}
\usepackage{epsfig}

\begin{document}

\def\Huge{\huge}
\def\e{\begin{equation}}
\def\f{\end{equation}}
\def\*{^{\displaystyle*}}
\def\xx{\displaystyle{{}^\times}\llap{${}_\times$}}
\def\x.{\displaystyle{{}^\times}\llap{${}_\cdot$}}
\def\.x{\,\displaystyle{{}^\cdot}\,\llap{${}_\times$}}
\def\=#1{\overline{\overline #1}}
\def\d{\partial}
\def\s{\strut\displaystyle}
\def\dz{{\partial\over\partial z}}
\def\Dz{{\partial\over\partial z\*}}
\def\u{\breve}
\def\df{{\partial\over\partial f}}
\def\Df{{\partial\over\partial f\*}}
\def\dt{{\partial\over\partial t}}
\def\-#1{\bar #1}
\def\o{\omega}
\def\E{\epsilon}
\def\M{\mu}
\def\D{\nabla}
\def\.{\cdot}
\def\x{\times}
\def\##1{{\bf#1\mit}}
\def\l#1{\label{eq:#1}}
\def\r#1{(\ref{eq:#1})}
\def\am{\left(\begin{array}{c}}
\def\amm{\left(\begin{array}{cc}}
\def\a{\end{array}\right)}
\newcommand{\ds}{\displaystyle}

\title{Electromagnetic wave refraction at an interface of a double wire medium}

  \author{Igor S. Nefedov$^{1}$, Ari J. Viitanen${^2}$, and
  Sergei A. Tretyakov${^1}$}

   \affiliation{$^{1}$Radio Laboratory / SMARAD, Department of Electrical and Communications
 Engineering,
 Helsinki University of Technology (TKK), P.O. Box 3000, FI-02015 TKK,
 Finland}

  \affiliation{$^{2}$Electromagnetics Laboratory, Department of Electrical and Communications
 Engineering, TKK} 

 \begin{abstract}
\noindent Plane-wave reflection and refraction at an interface
with a
 double  wire medium is considered. The problem of additional boundary
 conditions (ABC) in application to wire media is discussed and an ABC-free
 approach, known in the solid state physics, is used. Expressions for the
 fields and Poynting vectors of the refracted waves are derived.
 Directions and values of the power density flow of the refracted waves are
found and the conservation of the power flow through the interface is checked.
 The difference between the results, given
 by the conventional model of wire media and the model, properly taking into account spatial
 dispersion,  is discussed.
\end{abstract}

\maketitle

\section{Introduction}

Wire
medium (WM) is an artificial medium formed by a lattice of ideally
conducting thin wires. Recently, we have observed a growing interest to
such artificial media of possible new physical phenomena and potential applications.
This medium at low frequencies is usually described as a
uniaxial crystal, whose permittivity tensor components are expressed
by the plasma model. It has been shown \cite{SpatD}, that if the
wavevector in a WM has a nonzero component along the wires, the
plasma model should be corrected introducing spatial dispersion (SD).
Similarly, spatial dispersion is inherent to the double WM
(DWM) formed by two mutually orthogonal lattices of thin ideally
conducting straight wires, see Fig.~\ref{2DWM}.

\begin{figure}
\centering
 \epsfig{file=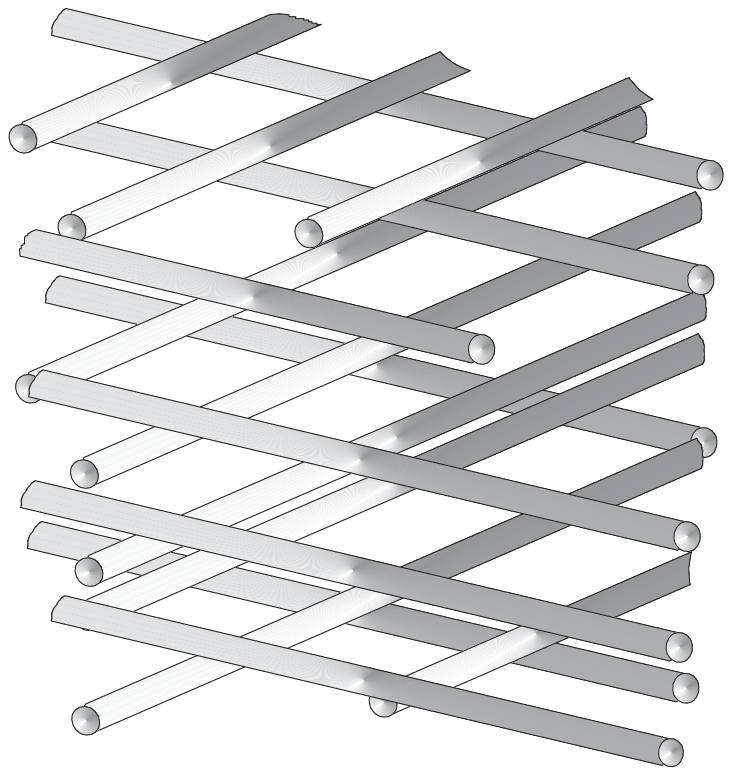, width=5cm}
 \epsfig{file=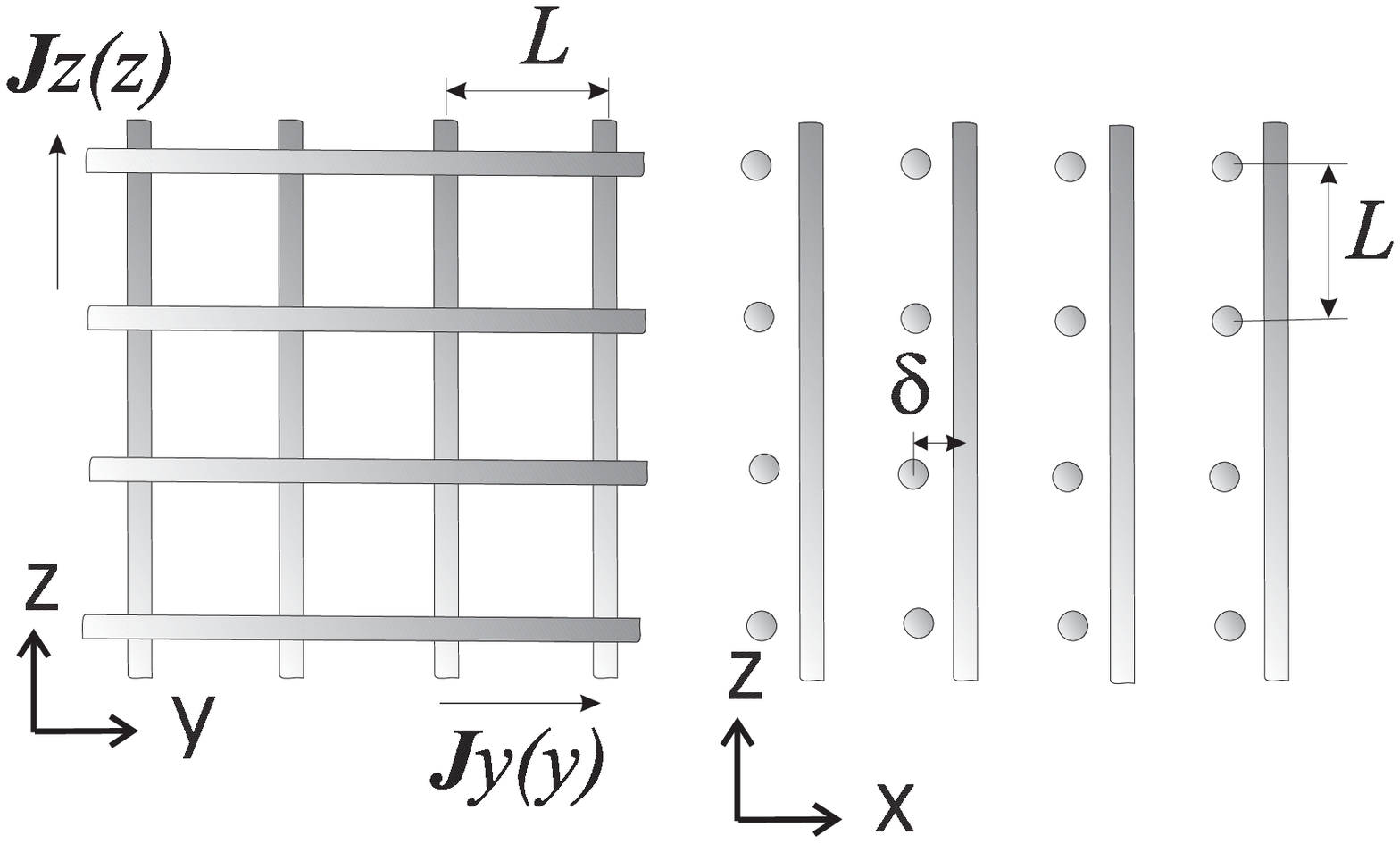, width=8cm}
 \caption{Geometry of the unbounded DWM.}
 \label{2DWM}
 \end{figure}

 For
consideration of waves in double wire media let us take a case where the
wires are perpendicular to each other in the $y$ and $z$ directions. The
waves in unbounded space filled with a DWM medium were studied in
\cite{Mario} numerically, in \cite{Belov} using a semi-analytical
approximation of the local field, and in \cite{Nefedov} both
numerically and using the effective medium (EM) approach. In the
last paper a very good agreement between the results given by the EM and
full-wave theories for all types of waves in DWM (if the wires
are thin) has been demonstrated.

In this paper we consider the plane-wave reflection and refraction
at an interface of DWM using the effective medium approach. We
assume that the two orthogonal wire arrays are identical, the period of the lattice
is equal to $L$ in the $x$, $y$, and $z$ directions, and the radius
of the wires is equal to $r_0$. In this case the wire lattice is
square in the plane of the wires, i.e., the $(yz)$ plane. We assume
also that the interface of DWM lies in the $(xy)$ plane and the
incident wave vector lies in the $(yz)$ plane (see Fig.~\ref{2DW}).
\begin{figure}[h]
 \centering
 \epsfig{file=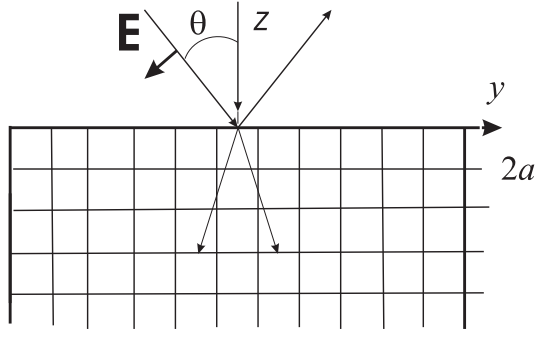, width=7cm}
  \caption{Geometry of the wave reflection problem.}
 \label{2DW}
 \end{figure}

In the next sections we will give basic expressions, obtained in the
framework of the EM approach, and formulate the wave refraction
problem, demonstrating necessity of additional boundary conditions
(ABC's) both for interface problems with single and double wire
media. Then we will discuss some approaches, applying in solid state
physics in order to overcome the ABC problem and look what may be
useful for us in application to WM. Using the ABC-free approach we
will find the reflection coefficient, amplitudes and Poynting
vectors for refracted waves. We will compare directions of the group
velocity and the energy  density flow found from the expression for
the Poynting vector containing additional terms inherent for media
with spatial dispersion. Conservation of the normal component of the
power flow vector at passing through the interface is checked.

\section{Field equations and eigenwaves in double unbounded wire
medium}

Assuming space-time dependence of fields as $e^{i(\omega
t-k_y y-k_z z)}$, there are non-zero wave vector components parallel to
wires.
  Anisotropy appears in this electromagnetic crystal with square lattice
  and DWM behaves as a biaxial crystal with the relative permittivity dyadic
  \e \=\epsilon=\epsilon_x\#u_x\#u_x+\epsilon_y\#u_y\#u_y+
\epsilon_z\#u_z\#u_z, \f
 with
  \e\l{a5}
\epsilon_x=\epsilon_h,\ \ \ \
\epsilon_y=\epsilon_h\left(1-{k_p^2\over{k^2-k_y^2}}\right),\ \ \ \
\ \epsilon_z=\epsilon_h\left(1-{k_p^2\over{k^2-k_z^2}}\right), \f
 where $k=\frac{\omega}{c}\sqrt{\E_h}$, $c$ is the speed of light,
and $\E_h$ is the relative permittivity of the host medium. Note, that
model \r{a5} works both for real and imaginary $k_y$ (for
propagating and evanescent waves, respectively), see \cite{SpatD}.

In the wire medium the Maxwell equations
 \e \nabla\x\#E=-j\omega
\mu_o\#H \f \e \nabla\x\#H=j\omega \epsilon_o\=\epsilon\.\#E \f
 split into two separate subsystems describing wave propagation of fields with
 TE and TM polarizations:
 \e
-j(k_y\#u_y+k_z\#u_z)\x(E_x\#u_x+E_y\#u_y+E_z\#u_z)=-j\omega \mu_o
(H_x\#u_x + H_y\#u_y+H_z\#u_z) \f \e
-j(k_y\#u_y+k_z\#u_z)\x(H_x\#u_x+H_y\#u_y+H_z\#u_z)=j\omega
\epsilon_o (\epsilon_x E_x\#u_x +
\epsilon_yE_y\#u_y+\epsilon_zE_z\#u_z). \f
 For ordinary (TE) waves this leads to the wave equation
 \e [k^2-k_y^2-k_z^2]E_x=0. \f
  The same
equations hold for $H_y$ and $H_z$. There are no effects due to
wires, and ordinary waves propagate as in any  isotropic dielectric
medium.

Whereas for extraordinary (TM) waves the wires affect the
propagation, and we obtain the wave equation
 \e\l{u1}
[k^2\epsilon_y-k_z^2 -k_y^2{\epsilon_y\over{\epsilon_z}} ]H_x=0. \f
 The same equation can be written for $E_y$ and $E_z$.

In order to solve the wave reflection problem we need to evaluate
the eigenwaves which are outgoing from the interface of DWM. It
means that we have to find $k_z$ under fixed $k$ and $k_y$. Let
$k_y=k\sin{\theta}$, where $\theta$ is the incidence angle.

It follows from \r{u1} that the dispersion equation has the form
\e\l{u2} T(k_z,\omega)=k_z^2- \left[k^2 \epsilon_y(k_y)
-k_y^2{\epsilon_y(k_y)\over{\epsilon_z(k_z)}}\right]=0,\f
 and its solution is:
  \e\l{u3}
 k_{z\pm}^2 =
 \frac{2k^4-2k^2k_p^2-3k^2k_y^2+2k_p^2k_y^2+k_y^4\pm
 k_y\sqrt{(k_y^2-k^2)((2k_p^2+k_y^2)^2-k^2(4k_p^2+k_y^2))}}{2(k^2-k_y^2)}.
 \f
Two waves propagating or attenuating in both directions follow from
the effective medium theory \r{u3}.
 The conventional isotropic
plasma model leads to only two waves for a certain direction, namely,
$k_z=\pm\sqrt{k_0^2\E-k_y^2}$, $\E=\E_h(1-k_p^2/k^2)$, where $k_0=\omega/c$.

For the following consideration of the refraction problem we will need
to know properties of waves propagating in the
$z$ direction. Here we briefly revise these properties (see our
previous paper \cite{Nefedov} for more details). The
real and imaginary parts of $k_z$ versus the normalized frequency
$k/k_p$ are presented in Fig.~\ref{CW} for $\theta=\pi/4$. Let us
assume that $\E_h=1$. In the simple case of the conventional model
(dotted curve) $k_z$ is imaginary if $k<K_2=k_p/\cos{\theta}$, and
it is real if $k>K_2$. The solution of the conventional model is
completely wrong between $K_1$ and $K_2$ because it gives an
imaginary value of $k_z$ instead of a real one which is obtained
from the correct model. The correct,
 more complicated solutions, follow from
Eq.~\r{u3}. Analyzing Eq.~\r{u3}, one can see that there exist
three frequency regions, corresponding to different kinds of
solution.

The first one is the low frequency band $k<K_1$, where
  \e\label{c1}
 K_1=k_p\frac{\sqrt{2}}{\sin{\theta}}\sqrt{\frac{1-\cos{\theta}}{\cos{\theta}}}
  \f
  is the stop band edge: Beyond this wave number the waves are propagating.
 There the propagation constant $k_z$ is complex despite the fact that we have assumed the
medium to be lossless
   (see Fig.~\ref{CW}). Actually, there are
 two complex conjugate solutions for each Re$(k_z)>0$.

The second frequency area is $K_1<k<K_2$. At point $K_2$ one of the
solutions is zero and within the range $K_1<k<K_2$ we have a forward
wave and a backward wave with respect to the interface. It means
that one of the waves has a positive projection of the wave vector
on the interface inner normal, while the other wave  has a negative
projection.

Finally, for $k>K_2$ both of the waves are propagating forward
waves. Electrodynamical calculations \cite{Nefedov} (using the three
dimensional Green's function) confirm the results of the effective
medium theory with a high accuracy in a wide spectral range
including the regions of evanescent and propagating waves (see solid
and dashed curves in Fig.~\ref{CW}). Note that point $K_2$
corresponds to the edge of the passband in the framework of the old
model. Thus, the model taking into account spatial dispersion leads
to a considerably more complicated structure of eigenwaves than the
conventional model of an isotropic plasma, and it is in very good
agreement with the results of the full-wave analysis.

\begin{figure}
\centering
 \epsfig{file=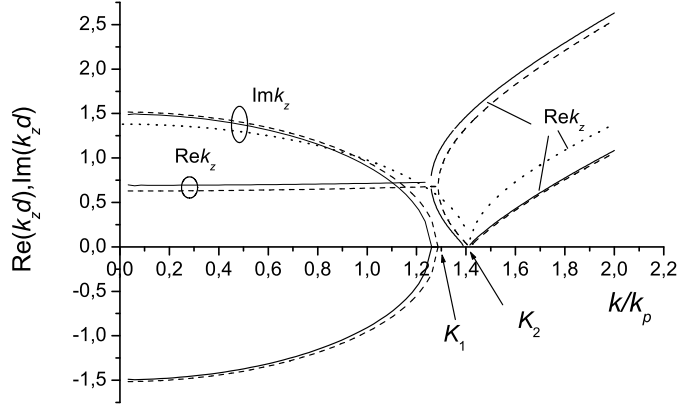, width=10cm}
 \caption{Real and imaginary parts of $k_z$,
 calculated using the electrodynamical model (solid curves) and the EM theory
(dashed curves).
 The dotted curve shows $k_z$ given by the conventional plasma model. }
 \label{CW}
 \end{figure}

\section{Wave reflection from a wire medium interface and the problem of
 additional boundary conditions}

As it was shown above,
  there exist two extraordinary waves
with the wave vector and the electric field in the $(yz)$ plane.
Assuming the $y$-component of the electric field of the incident
wave to be equal to unity, and applying the continuity conditions of
the tangential field components results in the reflection problem
formulated as follows:
\begin{equation}\label{q1}
 \begin{array}{l}
 1+R_E=E_++E_- \\
 (1-R_E)/Z_0=E_+/Z_++E_-/Z_-,
 \end{array}
 \end{equation}
where $R_E$ is the unknown reflection coefficient for electric field,
$E_+,\,E_-$ are the unknown
amplitudes of refracted waves in the wire medium, $Z_0$
is the wave impedance (TM) in free space
and $Z_{\pm}$
are the wave impedances of the refracted waves.
%
Thus the problem becomes similar to one appearing in
 crystallooptics, where excitons arise and spatial
   dispersion cannot be neglected \cite{Ginzburg}. The main
   difficulty here is the necessity to invoke additional boundary
   conditions (ABC) in order to match solutions at the interface
   of media. It was pointed out first by S.~Pekar \cite{Pekar} (1956), that the
   well-known Maxwell's boundary conditions (\ref{q1}) are not sufficient
   to connect the amplitudes of the incident and transmitted waves
   in adjoining media, if more than one independent wave can
   propagate in any medium.

Probably, the first ABC were proposed by S.~Pekar \cite{Pekar}, and
his ABC stay among the more often used in the theory of media with
SD. The simplest phenomenological Pekar's ABC are the following:
\begin{equation}\label{q2}
 \begin{array}{l}
{\bf P}=0, \\
 \frac{\d{\bf P}}{\d z}=0
 \end{array}
 \end{equation}
at the boundary, where {\bf P} is the polarization vector.
 Let us consider if we can apply to the WM the first condition of
 (\ref{q2}) in the region of evanescent waves.
  For the $y$-component of the polarization vector one
 can write
 \begin{equation}\label{q3}
 P_y=\chi_{y1}E_1+\chi_{y2}E_2=0,
 \end{equation}
 where the susceptibilities $\chi_{y1}$ and $\chi_{y2}$ relate to different waves,
but $\chi_{y1}=\chi_{y2}$. Hence we come to $E_1+E_2=0$, which means
that the reflection coefficient for the electric field $R_E=-1$. In
other words, a semi-infinite WM behaves as a perfect metal plane.
However, we know that it is not so even if only evanescent waves are
excited in the WM. The reflected wave changes its phase due to
penetration into WM, and this phase is not equal to $\pi$. The same
result is obtained considering condition $P_z=0$.
 The second Pekar's
condition (\ref{q2}) (for derivatives) leads to a similar result.
Thus, the conventional Pekar's ABC are useless for solving our
problem.

After pioneering Pekar's publication different ABC were proposed for
problems of crystallooptics, as well as semiconductor and plasma
electrodynamics. All of these works relate to specific media and
take into account the properties of a sub-surface layer at the media
interface (see \cite{Ginzburg,Halevi} and the bibliography
presented there). Besides, phenomenological assumptions and
experimental data are used in these theories. Since WM strongly
differs from a solid-state crystal, only general approaches to ABC
which do not concerned with particular media, may be interesting for
us.

 Many authors (see, for example, \cite{Hopfield,Zeyher})
derived ABC from a given model of medium with an explicit
specification of its surface. After simplifications these models
give nothing more than Pekar's ABC. Among different theories of SD
media so-called ``ABC-Free" theories attract our attention. One of
the ways for ABC derivation is a concept of "exciton dead layer",
proposed by Hopfield and Thomas. Actually, there is a layer in which
exciton wave function has evanescent form.
 Kikuo Cho \cite{Cho} has pointed out that there is a certain limit
of transition layer thickness, below which the ABC theory is
unnecessary. However, Cho theory is not yet free from some
parameters determined by specific surfaces. B.~Chen and D.F.~Nelson
declare that their work \cite{Nelson} solves the macroscopic ABC
problem completely. Despite that the authors of \cite{Nelson} used a
complicated quantum mechanical model, their derivation found that
the fully macroscopic solution is equivalent to the use of Pekar's
ABC $P(z=0)=0$. Thus, such an approach also is not suitable for us.
A.~Vinogradov and co-authors, considering in \cite{Alex} effects of
 SD in composite metamaterials, have obtained ABCs  using
an assumption of existence of additional waves in free space  which are
the same as in the metamaterial but have evanescent nature. This
assumption leads to additional equations at the interface.

   Another way to solve the problem was
proposed recently by
K. Henneberger \cite{Henn}.  It is
   based on the assumption of an abrupt transition from medium to vacuum.
 It is assumed that the incident wave excites a source $s(z,\omega)$, located within
a sub-surface layer $0<z<2a$, and its thickness is
assumed to be negligible. This approach is appropriate for our problem of
wire media interface. Indeed, it is known for problems of diffraction by
single semi-infinite wire grids that the induced currents deviate from the
regular amplitudes far from the interface only in a very narrow region whose
width is of the order of the grating period \cite{old}.
This conclusion holds for grids of wires both parallel and perpendicular to the
edge. For this reason we assume that for the wire medium interface the transition layer
has the thickness of only a few periods of the lattice.
This thickness is much smaller than the wavelength and negligible as compared with the
length of the wires (wires are infinite in our model).

Applying this approach to our interface problem of free space and
the wire medium, the wave equation for $H_x$ in unbounded medium
(Eq.~\r{u1} written for $H_x$) should be replaced by an
inhomogeneous equation
  \e\l{u4} {\partial^2 H_x\over{\partial
z^2}}+\left [k_0^2 \epsilon_y(k_y)
-k_y^2{\epsilon_y(k_y)\over{\epsilon_z(k_z)}}\right ]H_x=s(z,\omega),
 \f
 where $H_x$ is the refracted field.
  It means that any
 propagating wave has to be created by a source. The proper source of the
 penetrating wave in the wire medium is the incident wave and the
 polarization induced by it in the medium.
 Such an externally controlled source can be identified with some
 polarization additionally induced to the one already described
 by $\=\epsilon$.
 Therefore it is located only on the surface and in the transition
 region, where the induced polarization deviates from that in the
 bulk medium.
After Fourier transform of Eq.~\r{u4} one obtains
 \e\label{a10}
H_x(z,\omega) =\int_{-\infty}^{\infty} {\frac{d q}{2\pi}
\frac{s(q,\omega)e^{iqz}}{T(q,\omega)}},
   \f
where $s(q,\omega)$ is the Fourier transform of $s(z,\omega)$ and
$T(q,\omega)$ is determined by Eq.~\r{u2}. Assuming an abrupt
transition from the medium to vacuum, we can present the source as a
delta function $s(z,\omega)=s_0(\omega)\delta(z)$, then
$s(q,\omega)=s_0(\omega)$.
 If $T(q,\omega)$ is an analytical function, the integration in
 Eq.~(\ref{a10}) can be performed using the residue method.
Residues can be found by presenting $1/T$ in the form
  \e\l{u5}
{1\over{T(k_z,\omega)}}={1\over{k_z^2- \left[k_0^2 \epsilon_y(k_y)
-k_y^2{\epsilon_y(k_y)\over{\epsilon_z(k_z)}}\right ]}}=
{\beta_+\over{k_z^2-k_{z+}^2}}+{\beta_-\over{k_z^2-k_{z-}^2}}, \f
  where the coefficients are
   \e \beta_+={k^2-k_p^2-k_{z+}^2\over{k_{z-}^2-k_{z+}^2}} \f
\e \beta_-=-{k^2-k_p^2-k_{z-}^2\over{k_{z-}^2-k_{z+}^2}}. \f
 The residues (the relative amplitudes of the transmitted field
components) read
 \e
R_+={\beta_+\over{2k_{z+}}}=
{k^2-k_p^2-k_{z+}^2\over{2(k_{z-}^2-k_{z+}^2)k_{z+} }} \f \e
R_-={\beta_-\over{2k_{z-}}} =-{k^2-k_p^2-k_{z-}^2
\over{2(k_{z-}^2-k_{z+}^2)k_{z-} }}. \f

 Finally, the field component
$H_x$ in the wire medium is
 \e H_x
=s_o\left[R_+e^{-jk_{z+}z}+ R_-e^{-jk_{z-}z}\right ]
=s_o\left[{\beta_+\over{2k_{z+}}}e^{-jk_{z+}z}+
{\beta_-\over{2k_{z-}}}e^{-jk_{z-}z}\right ]. \f
 The expression for
$E_y$ and $E_z$ can be obtained from the Maxwell equations as
 \e
k_zH_x = -k_0\epsilon_y E_y, \ \ \ \ -k_yH_x = -k_0\epsilon_z E_z,
\f
 which gives us the electric field components in the wire medium
\e\l{v1} E_y =-{s_o\over{k_0\epsilon_y}} \left[\beta_+
e^{-jk_{z+}z}+ \beta_- e^{-jk_{z-}z}\right ] \f \e\l{v2} E_z= {k_y
s_o\over{\omega\epsilon_o}}
\left[{\beta_+\over{2k_{z+}\epsilon_{z+}}} e^{-jk_{z+}z}+
{\beta_-\over{2k_{z-}\epsilon_{z-}}} e^{-jk_{z-}z}\right].
 \f
  Now we have expressions for all field components induced in the wire medium
for the TM polarization.

In free space there exist incident and reflected TM waves. Magnetic
field components are
 \e H_x^i=H_0e^{-jk_yy}e^{-j\beta_0 z},\ \ \ \
H_x^r=H_re^{-jk_yy}e^{j\beta_0 z}=R_{H}H_0e^{-jk_yy}e^{j\beta_0 z},
\f
 where $\beta_0=\sqrt{k_0^2-k_y^2}$ and the electric field
components are
 \e
E_y^i=-{\beta_0\over{k_0}}H_0e^{-jk_yy}e^{-j\beta_0 z}, \ \ \ \
E_y^r= R_{H}{\beta_0\over{k_0}}H_0e^{-jk_yy}e^{j\beta_0 z} \f \e
E_z^i={k_y\over{k_0}}H_0e^{-jk_yy}e^{-j\beta_0 z}, \ \ \ \ E_z^r=
R_{H}{k_y\over{k_0}}H_0e^{-jk_yy}e^{j\beta_0 z}. \f
 At the interface
$z=0$ the continuity of the tangential field components leads to
relations
 \e H_0+R_{H}H_0={s_0\over{2}}\left[{\beta_+\over{k_{z+}}}
+{\beta_-\over{k_{z-}}}\right]\f \e
-{\beta_0\over{k_0}}H_0+{\beta_0\over{k_0}}R_{H}H_0 =
-{s_0\over{2k_0\epsilon_y}}\left[\beta_++\beta_-\right], \f
 from
which the reflection coefficient for magnetic field is obtained:
 \e
R_{H}= { \left({\beta_+\over{k_{z+}}}+{\beta_-\over{k_{z-}}}\right)
- {1\over{\epsilon_y}}\left({\beta_+\over{\beta_0}}
+{\beta_-\over{\beta_0}}\right) \over{
\left({\beta_+\over{k_{z+}}}+{\beta_-\over{k_{z-}}}\right) +
{1\over{\epsilon_y}}\left({\beta_+\over{\beta_0}}
+{\beta_-\over{\beta_0}}\right) }}. \f
  Finally, the explicit
expression for the coefficient $s_0$ (the transmission source) is
obtained:
 \e s_0={1+R_H\over{{1\over{2}}
({\beta_+\over{k_{z+}}}+{\beta_-\over{k_{z-}}})}}\ H_0={4H_0\over{
\left({\beta_+\over{k_{z+}}}+{\beta_-\over{k_{z-}}}\right) +
{1\over{\epsilon_y}}\left({\beta_+\over{\beta_0}}
+{\beta_-\over{\beta_0}}\right)}}. \f

In the region of complex waves $k/k_p<K_1$ we have to chose the
branches of the square roots
 for $k_{z\pm}$ having positive imaginary parts.
 In the region $K_1<k/k_p<K_2$ it is necessary to take for the
 backward wave ($``-"$ wave) the root branch with the opposite sign.
In the above derivation we evaluated the reflection coefficient for
the magnetic field. The reflection coefficient for the electric
field is $R_E=-R_H\cos{\theta}$.

\subsection{Reflection from a single wire medium interface}

The plane wave reflection coefficient from an interface of a single
wire medium where wires are along the $z$ axis is easily obtained as
a special case of the previously considered double wire medium
reflection problem. In a single wire medium the permittivity
components are $\epsilon_x=\epsilon_y=\epsilon_h$ and
$\epsilon_z=\epsilon_h\left(1-{k_p^2\over{k^2-k_z^2}}\right)$.
Evaluating the dispersion equation we have as solutions $k_{z+}=k$,
which is the propagation factor for the TEM mode and
$k_{z-}=\sqrt{k^2-k_y^2-k_p^2}$ for the TM mode. Thus, in the single
wire medium the two extraordinary eigenwaves are TEM and TM
polarized.

We can use exactly the same expressions for the reflection
coefficient as in the case of the double wire medium simply
substituting $k_{z+}=k$ and $k_{z-}=\sqrt{k^2-k_y^2-k_p^2}$. This
leads to expressions for the coefficients
 \e \beta_+={k_p^2\over{k_y^2+k_p^2}} \f
\e \beta_-={k_y^2\over{k_y^2+k_p^2}} \f
  and residues
   \e R_{+}={k_p^2\over{2k(k_y^2+k_p^2)}} \f
\e R_{-}={k_y^2\over{2\sqrt{k^2-k_y^2-k_p^2}(k_y^2+k_p^2)}}. \f

\section{Discussion}

 Fig.~\ref{R5} shows the phase of the reflected wave
 $\phi=\arctan{\left\{{\rm Im}(R_H)/{\rm Re}(R_H)\right\}}$ versus
 the
 incidence angle $\theta$, calculated for different $k/k_p$,
 corresponding to the region of complex waves (see Fig.~\ref{CW}).
Here Re$(R_H)$ and Im$(R_H)$ are the real and imaginary
 parts of the reflection coefficient $R_H$. The calculations have been
 performed using the model \r{a5}, taking into account spatial
 dispersion, and the conventional one,
 \e\label{aa5}
 \epsilon_y=\epsilon_z=\epsilon_h\left(1-\frac{k_p^2}{k^2}\right).
\f As it is expected, both  models give the modulus of the
reflection coefficient equal to unity due to the absence of losses
and propagating waves in wire
 medium. It is remarkable, that the spatial dispersion results in a weak dependence
 of the phase of the reflected wave on the incidence angle.
 The parameters of the wire medium are taken as above in the
 eigenvalues calculations.

  The real and imaginary parts of the reflection coefficient in a wide
  spectral range cover areas with complex waves, FW and BW waves
and FW waves only, are presented in Fig.~\ref{R6}. The incidence
angle is taken to be $\pi/4$. For
  understanding of these characteristics it is useful to compare them
  with the eigenvalue dispersion (Fig.~\ref{CW}). Distinctive
  points $K_1$, where propagating waves appear in the framework of new
  model, and $K_2$, where both of the waves become forward ones.
  Results, given by the old model, are also shown here. The most
  important feature is that the area of propagating waves shifts by $\Delta
  k/k_p=K_2-K_1$ in comparison with that given by the old model.

\begin{figure}[h]
 \centering
 \epsfig{file=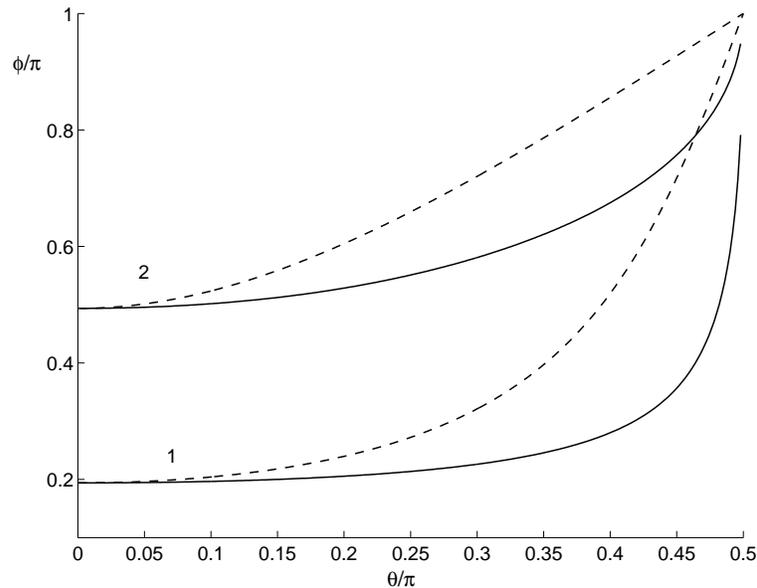, width=10cm}
\caption{Normalized phase of the reflected wave $\phi/\pi$ versus
the incidence angle $\theta/\pi$, calculated at different $k/k_p$:
curves~1 correspond to $k/k_p$=0.3 and curves~2 correspond to
$k/k_p$=0.7.
  Solid and dashed curves show the reflection phase given by the new and conventional
  models, respectively.}
 \label{R5}
 \end{figure}

\begin{figure}[h]
 \centering
 \epsfig{file=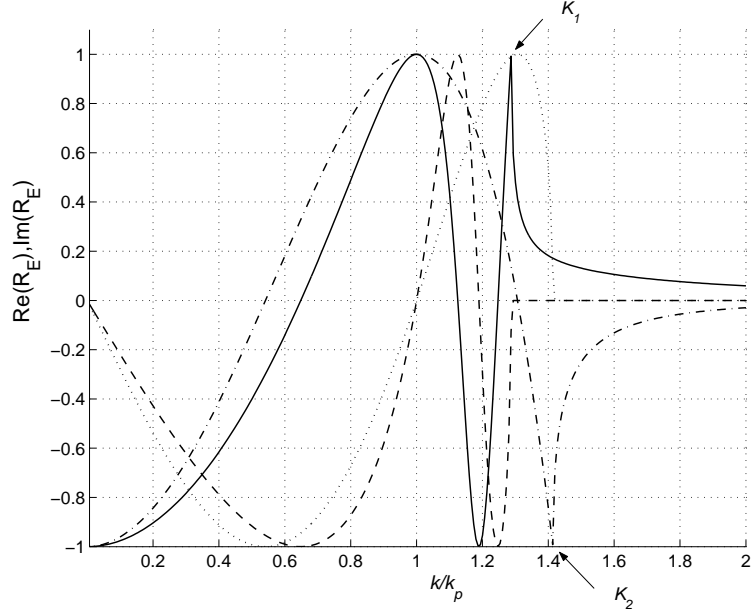, width=10cm}
\caption{Real and imaginary parts of the reflection coefficient
versus the normalized wavevector $k/k_p$. The solid line shows
Re$(R_E)$, the dashed line shows Im$(R_E)$ (the new model), the
dash-dot and dot lines correspond to Re$(R_E)$ and Im$(R_E)$,
respectively, obtained in the framework of the old model.
  }
 \label{R6}
 \end{figure}

 \section{Group velocity, Poynting vector and refracted waves in DWM}

In this section we will discuss the group velocity and Poynting
vectors of waves excited in a double wire medium and check the power
conservation at the interface.
 It is well known that the group velocity is defined as
 \e\l{u7}
 v_{gr}={\rm grad}_{\bf k}\omega. \f
 The Poynting vector that  determines the energy density flow  for media with
spatial dispersion has the form \cite{Landau}:
 \e
\#S={1\over{2}}{\rm Re}\{\#E\times\#H^*\}-{\omega\over{4}}{\partial
\epsilon_{ik}\over{\partial{\#k}}}E^*_iE_k, \f
 where the permittivity
dyadic components are expressed by formulae \r{a5} for DWM. Their
partial derivatives read
  \e {\partial \epsilon_{x}\over{\partial
k_x}}=0,\ \ \ {\partial \epsilon_{y}\over{\partial
k_y}}=-{2k_p^2k_y\over{(k^2-k_y^2)^2}},\ \ \ {\partial
\epsilon_{z}\over{\partial k_z}}=-{2k_p^2k_z\over{(k^2-k_z^2)^2}}.
\f

Let us derive expressions for the Poynting vector in free space and
in the wire medium. In free space the field expressions are
 \e \#H_1=H_0[e^{-j\beta_0 z}+R_H e^{j\beta_0 z}]e^{-jk_yy}\#u_x \f
\e \#E_1= {\beta_0H_0\over{k_0}} [-e^{-j\beta_0z}+R_He^{j\beta_0
z}]e^{-jk_yy}\#u_y + {k_yH_0\over{k_0}}
[e^{-j\beta_0z}+R_He^{j\beta_0 z}]e^{-jk_yy}\#u_z. \f
 The fields of
the  waves, marked by $+$ and $-$, in the wire medium are
 \e
\#H_{2\pm}={s_0\over{2}}{\beta_{\pm}\over{k_{z\pm}}}e^{-jk_{z\pm}z}
e^{-jk_yy}\#u_x, \f
 \e \#E_{2\pm}= -{s_0\eta\over{2k_0\epsilon_y}}
\beta_{\pm}e^{-jk_{z\pm}z}e^{-jk_yy}\#u_y +{k_ys_0\eta\over{2k_0}}
{\beta_{\pm}\over{k_{z\pm}\epsilon_{z\pm}}}e^{-jk_{z\pm}z}e^{-jk_yy}\#u_z.
  \f
 Now we can write the Poynting vector in free space and in the wire
medium (at the interface):
  \e \#S_1(0)={1\over{2}}{\rm Re}\{\#E_1\x\#H_1^*\}
= {|H_0|^2\eta\over{2k_0}}\left[\beta_0(1-|R_H|^2)\#u_z+
k_y(1+2|R_H|\cos{\phi}+|R_H|^2)\#u_y\right] \f
  with the notation
$R_H=|R_H|e^{j\phi}$. In the wire medium, the cross product term is
 \e\l{w1} \#S_{2\pm}^0(0)={1\over{2}}{\rm Re}\{\#E_{2\pm}\x\#H_{2\pm}^*\}
={|s_0|^2\eta\over{2k_0}}{1\over{4}}\ {\rm Re}\left\{
{1\over{\epsilon_y}}{\beta_{\pm}\beta_{\pm}^*\over{k_{z\pm}^*}}
\#u_z
+k_y{\beta_{\pm}\beta_{\pm}^*\over{k_{z\pm}k_{z\pm}^*\epsilon_{z\pm}}}
\#u_y \right\}, \f
  and the spatial dispersive term is
    \e
\#S_{2\pm}^d(0)={|s_0|^2\eta\over{2k_0}}{1\over{4}}\ \left[
{k_p^2k_y^2k_{z\pm}\beta_{\pm}\beta_{\pm}^*\over{(k^2-k_{z\pm}^2)^2
k_{z\pm}k_{z\pm}^*\epsilon_{z\pm}\epsilon_{z\pm}^*}} \#u_z
+{k_p^2k_y\over{(k^2-k_y^2)^2\epsilon_y^2}}
\beta_{\pm}\beta_{\pm}^*\#u_y \right]. \f
 The total Poynting vector in wire medium is
  \e\l{v4}
\#S_{2\pm}(0)=\#S_{2\pm}^0(0)+\#S_{2\pm}^d(0). \f

As an example, we consider excitation of modes, located above the
plasma frequency \cite{Mario}.
  For the taken parameters of DWM, the wire radius
 $r_0$=0.01~cm, and the period $L$=1~cm, the plasma wavenumber is $k_p\approx
 1.38$~cm$^{-1}$. Dispersion diagram in form of isofrequencies
and directions of the Poynting vector (the energy velocity) for
 these modes with respect to the normal to the interface ($z$-axis)
 are shown in Fig.~\ref{Group}. The calculations were performed at
 $k/k_p=1.35$. The value of the tangential component
$k_y$ is determined by the incidence angle $\theta$, namely,
$k_y=k\sin{\theta}$. The direction of the group velocity, found by
numerical differentiation of the dispersion characteristics, exactly
coincides with the direction of $\#S_{2\pm}$, calculated using formulae
\r{w1}--\r{v4}. Disregarding the term that takes into account spatial
dispersion leads to a strongly incorrect result, illustrated by dashed curves.

\begin{figure}[h!]
\centering
 \epsfig{file=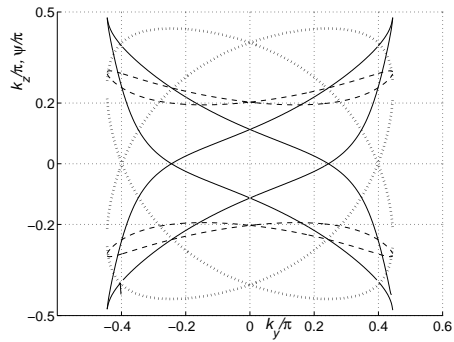, width=6cm}\caption{1. Isofrequencies, dotted curves.
2. Solid curves show the angle $\psi$ between the energy
velocity and the $z$ axis versus $k_y$. 3. Dashed curves show the
angle between $\#S_{2\pm}^0(0)$ and the $z$ axis versus $k_y$. }\label{Group}
\end{figure}

Next, let us consider the power conservation at the interface of
free space and the wire medium. It is important to check this
properly because in the wire medium region there exist two waves,
and the wire medium is spatially dispersive. In the frequency range
considered here the parameter values are assumed to be real. The
normal components of the Poynting vector on both sides of the
interface are (using the continuity condition of the tangential
field components)
  \e
S_{z1}(0)={|s_o|^2\over{8\omega\epsilon_o\epsilon_y}}
\left({\beta_+\over{k_{z+}}}+{\beta_-\over{k_{z-}}}\right), \f
  and
\e S_{2z}(0)={|s_o|^2\over{8\omega\epsilon_o}}\left[
{1\over{\epsilon_y}}\left({\beta_+^2\over{k_{z+}}}
+{\beta_-^2\over{k_{z-}}}\right)+
{k_p^2k_y^2\beta_+^2\over{(k^2-k_{z+}^2)^2 \epsilon_{z+}^2k_{z+}}} +
{k_p^2k_y^2\beta_-^2\over{(k^2-k_{z-}^2)^2 \epsilon_{z-}^2k_{z-}}}
\right]. \f
 Subtracting these power density components leads to
expression
 \e S_{z1}(0)-S_{z2}(0)= {|s_o|^2\over{8\omega\epsilon_o}}
\left({1\over{k_{z_+}}}+{1\over{k_{z_-}}}\right) \left[
{\beta_+\beta_-\over{\epsilon_y}}-
{k_p^2k_y^2\over{(k_{z-}^2-k_{z+}^2)^2}} \right]. \f
 Using the
expressions for $\beta_{\pm}$, $k_{z\pm}^2$ and $\epsilon_y$, we
find that the term inside the square brackets vanishes. The normal
component of the Poynting vector is continuous across the interface,
which means that the power conservation law is satisfied.

The normal to the interface components of the Poynting vectors  of
the refracted waves as well as the reflection coefficient (which is
purely real beyond $K_1$) are shown in Fig.~\ref{S1S2}. The values
of $S_{2z+}$ and $S_{2z-}$ are normalized to the power density of
the incident plane wave $|S_i|=\frac12\eta H_0^2\cos{\theta}$. These
results agree with those shown in Fig.~\ref{CW}, namely, $S_{2z-}$
becomes zero at point $k/k_p=K_2$. The point $K_2$ is a transition
point of the $``-"$ wave because it is a backward wave with respect
to the interface if $k/k_p<K_2$, and a forward wave if $k/k_p>K_2$.

\begin{figure}[h!]
\centering
 \epsfig{file=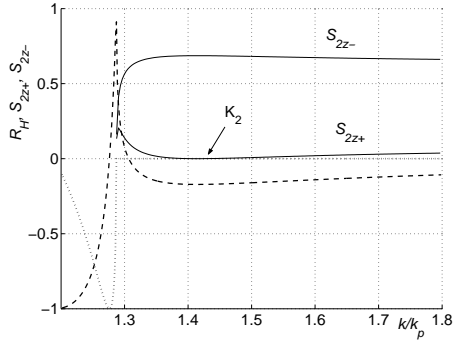, width=6cm}\caption{Reflection coefficient $R_H$ (dotted curve),
$S_{2z+}$ (solid curve) and $S_{2z-}$ (dashed curve) versus the
normalized wave vector.}\label{S1S2}
\end{figure}

\begin{figure}
\centering
 \epsfig{file=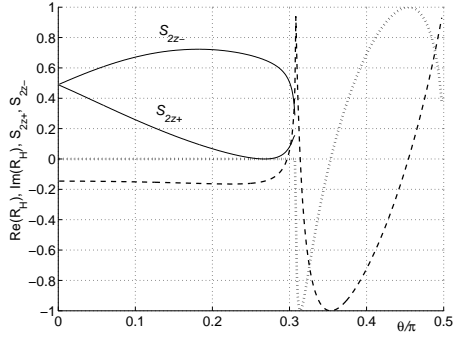, width=6cm}
\caption{The real and imaginary parts of $R_H$ (dashed and dotted
curves, respectively),
 $S_{z1}$ and $S_{z2}$ (solid curves) versus the incidence angle, calculated at $k/k_p=1.5$.}
 \label{A2d}
 \end{figure}

 Angular dependence of the normal components of the Poynting vector is
 presented in Fig.~\ref{A2d}.
Both refracted modes are TM modes in this case. At low frequencies
($k/k_p<1$) a plane wave cannot excite propagating TM modes at any
incidence angle $\theta$, so we present here the results for
$k/k_p=1.5$.
 There exists a critical angle $\theta_c\simeq 0.3\pi$, such that
$S_{z1}=S_{z2}=0$ if $\theta>\theta_c$. As in the previous case,
$R_H=+1$ if $\theta=\theta_c$, and $R_H\rightarrow 1$ if
$\theta\rightarrow\pi/2$, so DWM is an electric wall for grazing
incidence.

\subsection{Poynting vector in single wire media}

The Poynting vector in a single wire medium can be simply deduced
from the expressions for the double wire media. In this case the
partial derivatives of the permittivity components read
  \e {\partial
\epsilon_x\over{\partial k_x}}=0,\ \ \ {\partial
\epsilon_y\over{\partial k_y}}=0,\ \ \ {\partial
\epsilon_z\over{\partial k_z}}=-{2k_p^2k_z\over{(k^2-k_z^2)^2}}. \f
The additional term arises now only in the $z$ component of the
Poynting vector, and it is equal to
  \e
\#S_{2\pm}^d={|s_o|^2\eta\over{2k_o}}{1\over{4}}
{k_p^2k_y^2|\beta_{\pm}|^2\over{k_{z\pm}(k^2-k_{z\pm}^2-k_p^2)^2}}\#u_z.
\f
 Thus, the total Poynting vector reads for real parameter
values
 \e
\#S_{2\pm}={|s_o|^2\eta\over{2k_o}}{\beta_{\pm}^2\over{4k_{z\pm}^2}}\left[
{k_y\over{\epsilon_{z\pm}}}\#u_y+k_{z\pm}\left({1\over{\epsilon_y}}+{k_p^2k_y^2
\over{(k^2-k_{z\pm}^2-k_p^2)^2}}\right)\#u_z\right]. \f
  This gives us
after substituting $k_{z\pm}$ and $\beta_{\pm}$
 \e
\#S_{2+}={|s_o|^2\eta\over{2k_o}}{k k_p^2\over{4(k_y^2+k_p^2)}}\#u_z
\f and
 \e \#S_{2-}={|s_o|^2\eta\over{2k_o}}{k_y^2+k_p^2\over{4k_y^2}}
\left(k_y\#u_y+\sqrt{k^2-k_y^2-k_p^2}\#u_z\right). \f

 It is known
that for each eigenwave the group velocity in single wire media is
parallel to the phase velocity. From the above expressions one can
see that the Poynting vector is also in the same direction as the
group velocity (and the phase velocity). The continuity condition of
the power flow across the interface (continuity of the normal
component of the Poynting vector) can be easily checked similarly as
for double wire media.

\begin{figure}
\centering
 \epsfig{file=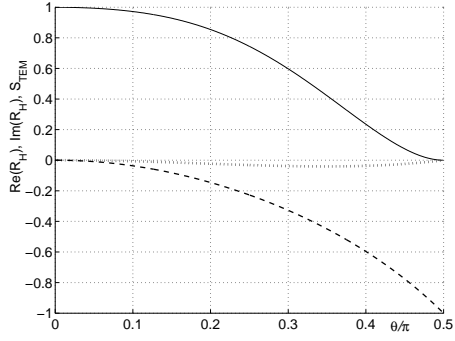, width=6cm}
 \caption{The real and imaginary parts of $R_H$ (dashed and dotted curves, respectively),
 $S_{TEM}$ (solid curves) versus the incidence angle, calculated at $k/k_p=0.5$.}
 \label{A1d05}
 \end{figure}

Fig.~\ref{A1d05} illustrates the angular dependence of the
reflection coefficient $R_H$ and the normal to the interface
component of the Poynting vector $S_z$ for the TEM mode ($S_{TEM}$).
Parameters of the WM are taken the same as in the case of DWM. The
amplitude of the magnetic field of the incident wave is $H_0=1$, and
the calculated Poynting vector is normalized to the incident wave
power flow density. Wavenumber $k/k_p$ corresponds to the frequency
below the plasma resonance and the TM mode cannot be excited at any
$\theta$. For grazing incidence $R_H\rightarrow -1$. It is obvious
that in this case the surface must behave as a wall, because no
traveling waves in the medium can be excited. Actually, it behaves
as a magnetic wall, because in this geometry the wires are
orthogonal to the interface, and the normal component of the
electric field is zero.
\begin{figure}
\centering
 \epsfig{file=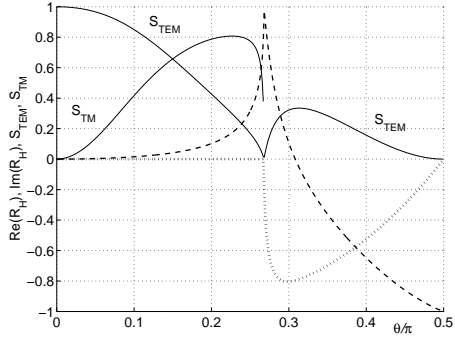, width=6cm}
 \caption{The real and imaginary parts of $R_H$ (dashed and dotted curves, respectively),
 $S_{TEM}$ and $S_{zTM}$ (solid curves) versus the incidence angle, calculated at $k/k_p=1.5$.}
 \label{A1d15}
 \end{figure}

More complicated is the case when the wavenumber is taken beyond the
plasma resonance, i.e. $k/k_p=1.5$, see Fig.~\ref{A1d15}.
 However, it follows from Eq.~\r{a5}, that for oblique
wave incidence, the ``effective" plasma wavenumber becomes
$k'_p=k_p/\cos{\theta}$ because $k_y=k\sin{\theta}$. For grazing
incidence $R_H\rightarrow -1$, similarly to the previous case.

For small incidence angles both TEM and TM modes are excited. There
exists an incidence angle $\theta'$, which is the angle of total
reflection.
 It takes place when $k=k'_p$, so $\cos{\theta'}$=2/3.
Neither TEM nor TM modes are excited, and $R_H\rightarrow 1$, which
corresponds to the case of an electric wall. For larger $\theta$ the
TM mode disappears.

\section{Conclusions}

Considering the problem of a plane wave refraction at the interface
of a double wire medium, exhibiting strong spatial dispersion, we
have shown, that Pekar's additional boundary conditions are not
applicable for its solution. We have analyzed the literature discussion
on the ABC problem and have come to the point of view, that
Hennenberger's approach \cite{Henn} for SD media can be applied for
some kinds of metamaterials in the microwave range including single
and double wire media.

 Despite of criticism of the used ABC-free method, see
 \cite{CN,CZ} and also replies \cite{HR}, we suppose that
application of this approach has allowed us to overcome the problem of
 additional waves appearing in media with spatial dispersion and
to obtain reasonable results for reflection coefficient and for the
power densities and group velocities of the refracted waves.
Fulfilment of the conservation of the power, passing through the
interface, can be considered as an evidence of the correctness of
this approach.

Possible applications of DWM include antenna structures,
low-frequency filters, frequency selective radomes, double negative
metamaterials.

\subsection*{Acknowledgements}

This work has been performed in the frame of the {\it Metamorphose} Network of
Excellence and partially funded by the Academy of Finland and TEKES
through the Center-of-Excellence program.


\begin{thebibliography}{99}


 \bibitem{SpatD}
P.A. Belov, R. Marques, S.I. Maslovski, I.S. Nefedov,
 M. Silveirinha, C.R.~Simovski, and S.A.~Tretyakov,
  Phys.~Rev.~B  {\bf~67}, 113103 (2003).
\bibitem{Mario}
M.G. Silveirinha, and C.A. Fernandes, IEEE Trans. Microwave Theory
Tech. {\bf~52}, 889 (2004).
\bibitem{Belov}
C.R. Simovski and P.A. Belov, Phys. Rev.~E {\bf 70}, 046616 (2004).
\bibitem{Nefedov}
I.S. Nefedov, A.J. Viitanen, S.A. Tretyakov. Phys. Rev. E {\bf ~71},
046612 (2005).
\bibitem{Silin}
  R.A. Silin, {\it Periodic Waveguides}, Phasis, Moscow, 2002.
  \bibitem{HN}
S. Tretyakov, I. Nefedov, A. Sihvola, S. Maslovski, and C. Simovski,
J.~of Electrom. Waves and Applic {\bf 17}, 595 (2003).
\bibitem{Ginzburg}
V.M. Agranovich, V.L. Ginzburg, {\it Crystal optics with spatial
dispersion and excitons} (2nd ed., New York: Springer, 1984).

\bibitem{Pekar}
S.I. Pekar, Zh. Eksp. Teor. Fiz. {\bf~33}, 1022 (1957). [Soviet
Phys. JETP {\bf~6}, 785 (1958)].

\bibitem{Halevi}
P. Halevi, {\it Spatial Dispersion in Solids and Plasmas}
(North-Holland. Amsterdam, 1992), p. 339. P.~Halevi, in {\it
Excitons in Confined Systems}, Ed. by R.~Del~Sole, A.~D'Andrea, and
A.~Lapiccirela, Springer Proceedings in Physics Vol.~25
(Springer-Verlag. Berlin, 1988), p.~2.

\bibitem{Hopfield}
J.J. Hopfield and D.G. Thomas, Phys. Rev. {\bf 132}, 563 (1963).
 \bibitem{Zeyher}
R. Zeyher, J.L. Birman, and W. Brenig, Phys. Rev.~B {\bf~6}, 4613
(1972).
 \bibitem{Cho}
 Kikuo Cho, J. of the Physical Society of Japan
{\bf~55}, 4113 (1986).
 \bibitem{Nelson}
  B. Chen and D.F. Nelson, Phys. Rev. B {\bf 48}, 15372 (1993).
\bibitem{Alex}
A.P. Vinogradov, A.A. Kalachev, A.N. Lagarkov, V.E. Romanenko, and
G.V. Kazantseva, Doklady Physics {\bf~41}, 291 (1996).

\bibitem{Henn}
K. Henneberger, Phys. Rev. Lett. {\bf~80}, 2889 (1998).

\bibitem{old}
V.A. Rozov, S.A. Tretyakov,
Radioengineering and Electronic Physics {\bf 29}, 37 (1984).
\bibitem{Landau}
L.D. Landau and E.M. Lifshitz, {\itshape Electrodynamics of
Continuous Media (Course of Theoretical Physics, Volume 8),}
Butterworth-Heinemann; 2 edition, 1984.

\bibitem{CN}
D.F. Nelson, B. Chen, Phys.~Rev. Lett. {\bf~83}, 1263 (1999).
\bibitem{CZ}
R. Zeyher, Phys.~Rev. Lett. {\bf~83}, 1264 (1999).
\bibitem{HR}
K. Henneberger, Phys. Rev. Lett. {\bf~83}, 1265 (1999).


 \end{thebibliography}
\end{document}